\documentclass{ws-p8-50x6-00}

\input{epsf}
\usepackage{amsfonts}

\def\Journal#1#2#3#4{{#1} {\bf #2}, #3 (#4)}

\def\YAD{\em Yad. Fiz.}
\def\NCA{\em Nuovo Cimento}

\def\PLB{{\em Phys. Lett.}  B}
\def\PRL{\em Phys. Rev. Lett.}
\def\PRD{{\em Phys. Rev.} D}
\def\ZPC{{\em Z. Phys.} C}
\def\EPG{{\em Eur.Phys.J.} C}

\def\be{\begin{equation}}
\def\ee{\end{equation}}
\def\bea{\begin{eqnarray}}
\def\eea{\end{eqnarray}}

\begin{document}

\title{REGGE-EIKONAL APPROACH AND ITS OFF-SHELL EXTENTION VERSUS
EXPERIMENTAL 
DATA}

\author{V. A. PETROV and \underline{A. V. PROKUDIN}}

\address{
\vskip 0.2cm 
Institute For High Energy Physics,\\ 
142284 Protvino , Russia\\E-mail: petrov@mx.ihep.su \\
prokudin@th1.ihep.su}
\vskip 0.2cm




\maketitle\abstracts{ We develop an off-shell extention of the
Regge-eikonal approach which automatically takes into account off-shell
unitarity. We argue that exclusive vector-meson production cross-sections
measured at HERA can be fairly described with classical universal Regge
trajectories. No extra ``hard'' trajectories of high intercept are needed
for that.}

\section{Introduction}
Unitarity condition

$$
Im T(s,\vec b) = \vert T(s,\vec b)\vert^2 + \eta (s,\vec b)
$$
\noindent
where $T(s,\vec b)$ is the scattering amplitude in impact space
representation and $\eta (s,\vec b)$ stands for the
contribution of 
inelastic channels, may be illustrated by the following picture in the
momentum space 

\begin{figure}[h]
\vskip -5cm
\hskip 1cm {\vbox to 50mm{\hbox to 50mm{\epsfxsize=120mm
\epsffile{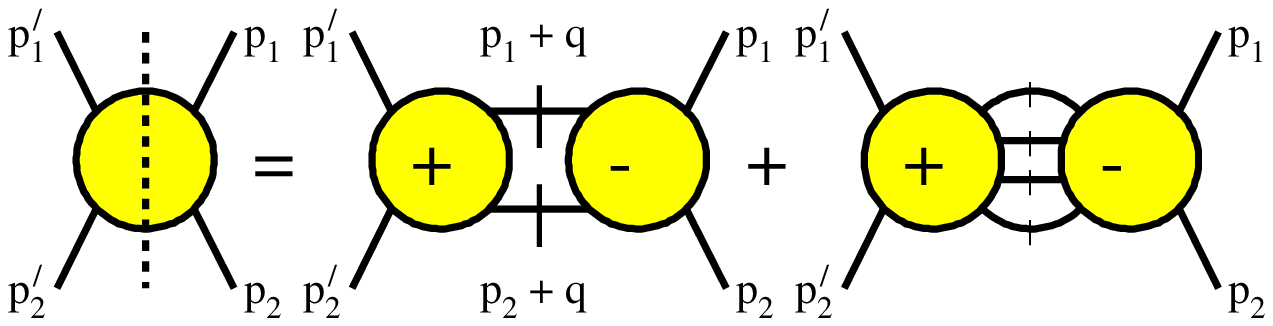}}}}
\vskip 4cm
\end{figure}

\be
ImT(p_{1}',p_{2}',p_1,p_2)=\frac{-i}{(2\pi)^4}\int d^4 q (-2\pi i)^2 \delta
((p_1 +q)^2 - m^2)
\delta ((p_2 -q)^2 - m^2)\times$$ \\ \vskip -6mm $$\times T^+
(p_1+q,p_2-q,p_{1}',p_{2}')T^- (p_1,p_2,p_{1}+q,p_{2}-q)+...
\label{eq:tt}
\ee
\noindent
where ``...'' stands for inelastic channels contribution.

It is worth noting that we can include virtual external particles into
consideration
and in this case the integration in Eq.~(\ref{eq:tt}) still goes over
particles
on mass shell as it was in the case of real external particles, i.e.

\begin{figure}[h]
\vskip -5cm
\hskip 1cm {\vbox to 50mm{\hbox to 50mm{\epsfxsize=120mm
\epsffile{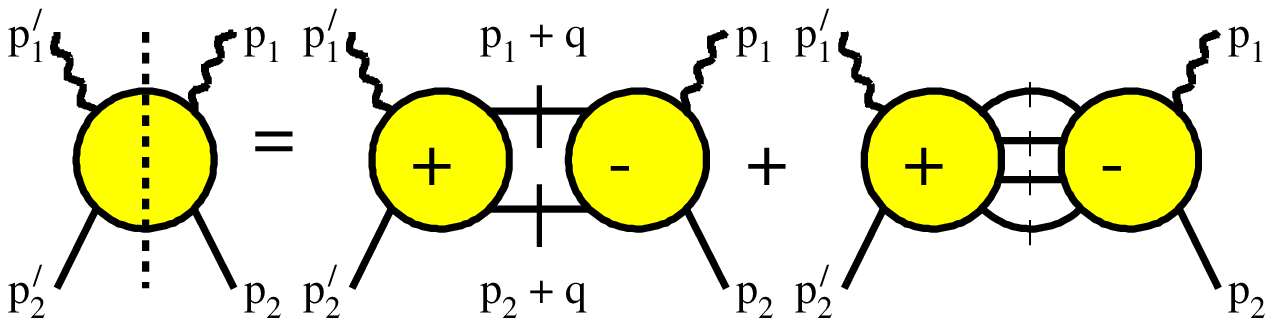}}}}
\vskip 4cm
\end{figure}

There are many parametrizations that ``solve" the $s$-channel unitarity
condition. Examples are $U$-matrix ~\cite{tyurin} and Eikonal approach
~\cite{re}.

We choose the eikonal representation as usual
\begin{equation}
T(s,\vec b)=\frac{e^{2i\delta (s,\vec b)}-1}{2i}\; ,
\label{eq:ampl}
\end{equation}
here $T(s,\vec b)$ is the scattering amplitude, $\delta (s,\vec b)$ is the
eikonal function. The eikonal function is to be considered as a basic
scattering
function, which builds up the amplitude. The unitarity condition looks very
simple in terms of the eikonal function:

\begin{equation}
Im \delta (s,\vec b) \ge 0, \; s>s_{inel}
\label{eq:euc}
\end{equation}

In quantum mechanics $\delta (s,\vec b)$ is related to the
(short-range)potential $V$:

\newpage

\begin{minipage}[b]{2.5cm}
$$
\delta (s,\vec b) \sim \int_{-\infty}^{+\infty}dz V(z,\vec b)\sim 
$$
\end{minipage}
\vskip -2cm
\begin{figure}[h]
\hskip 5cm {\vbox to 25mm{{
\epsfysize=25mm\epsffile{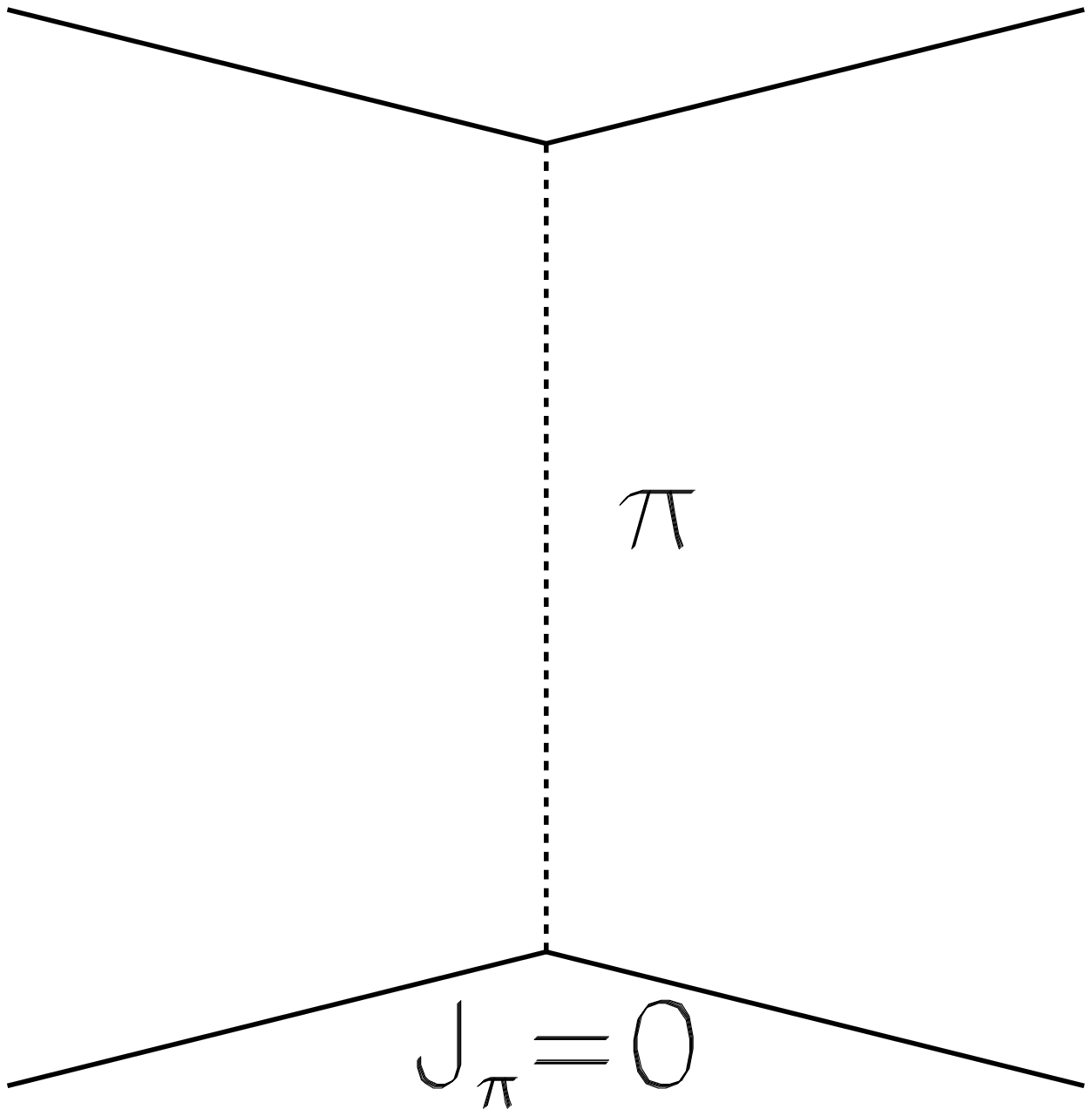}}}}
\end{figure}

At relativistic energies one can generalize the notion of the potential by
introducing a ``quasipotential'', which is complex and depends on energy
~\cite{qp}. As a concrete realizaton of the quasipotential one can take
Van-Hove interpretation of the Regge behaviour as a ``sum'' of all possible
one-particle exchanges in the $t$-channel ~\cite{vh}. Combination of the
eikonal approach and Regge poles was considered for the first time in
~\cite{re}. It was shown in ~\cite{cheng} in the framework of the
asymptotic summation of the perturbative series that the eikonal function
$\hat \delta (s, t)$ behaves like $s^{1+\Delta}$, where $\Delta > 0$. 

So one could abstract the idea that there exists some function $\hat \delta
(s, t)$ with a powerlike growth in $s$ and this is in full conformity with
general principles.

We choose the following eikonal function in $t$ space (here $t$ is
the transferred momentum)
\be
\hat \delta (s,t) = c \Big( \frac{s}{s_0} \Big)
^{\alpha(0)}e^{t\frac{\rho^2}{4}}
\label{eq:eikonalt}
\ee
where  
\be
\rho^2 = 4\alpha'(0) ln\frac{s}{s_0}+r^2
\ee
\noindent
is reffered to as a ``reggeon radius''.

It means that the eikonal function has a simple pole in $J$ plane and the
corresponding Regge trajectory can be written in the following form:
\be
\alpha(t) = \alpha(0) + \alpha '(0)t
\label{eq:rt}
\ee
We make use of linear Regge trajectories in this paper, though there exists
some evidence of non-linearity of $\alpha(t)$ when $t$ is high enough.

Functions in $t$- and $b$-spaces are related by the Fourier-Bessel
transformation:
\be
\begin{array}{r}
\hat f(t)= 4 \pi s\int_{0}^{\infty} db^2 J_0(b\sqrt{-t}) f(b) \\
\\
f(b)= \frac{1}{16 \pi s}\int_{-\infty}^{0} dt J_0(b\sqrt{-t}) \hat f(t) 
\end{array}
\label{eq:fb}
\ee

Making use of Eq.~(\ref{eq:fb}) we obtain the following $b$-representation
of the eikonal function:
\be
\delta (s,b) = \frac{c}{s_0}
\Big(\frac{s}{s_0}\Big)^{\alpha(0)-1}\frac{e^{-\frac{b^2}{\rho^2}}}{4\pi
\rho^2}
\label{eq:eikonalb}
\ee

In what follows the term ``pomeron'' will mean the leading
pole of the eikonal function.

For cross sections we use the following normalizations:
\be
\begin{array}{l}
\sigma_{tot} = \frac{1}{s} ImT(s,t=0) \\
\\
\sigma_{el} = 4 \pi \int_{0}^{\infty}db^2 \vert T(s,b) \vert^2 \\
\\
\frac{d\sigma}{dt} = \frac{\vert T(s,t) \vert^2}{16\pi s^2}
\end{array}
\label{eq:norm}
\ee

\section{Regge-Eikonal Approach}
Let's recall some properties of the Regge-eikonal model.

If $s\rightarrow \infty$, then 
\be
\sigma_{tot} \rightarrow 2 \pi \rho^2 [c+ln z-Ei(-z)]
\ee
where $z=\frac{\frac{c}{s_o}(\frac{s}{s_o})^\Delta}{2 \pi \rho^2}$, here
($\Delta \equiv \alpha (0)-1$) and $Ei(z)=
\int_{-\infty}^{z}\frac{e^x}{x}dx$.
At extra high energies we reach the following limit:
\be
\sigma_{tot} \rightarrow 8 \pi \alpha'(0)\Delta \Big(ln\frac{s}{s_0}\Big)^2
\ee
i.e. the powerlike behaviour of $\hat \delta (s,t)$ leads to the asymptotic
behaviour of the full scattering ampitude which functionally saturates the
Froissart-Martin bound~\cite{fm}:
\be
\sigma_{tot}^{hh} \le \frac{\pi}{m_{\pi}^2}\Big(ln\frac{s}{s_0}\Big)^2,
s\rightarrow \infty \;.
\ee

If $s\rightarrow \infty$, then
\be
\sigma_{el} \rightarrow  \pi \rho^2 [c+ln \frac{z}{2}+Ei(-2z)-2Ei(-z)]
\ee
and thus
\be
\frac{\sigma_{el}}{\sigma_{tot}} \rightarrow \frac{1}{2}
\label{eq:ratio}
\ee
\section{The Model}
In our model we choose the following contributions for the eikonal function:
\be
\delta(s,b) = \delta^+_{\Bbb P}(s,b)+ \delta^-_{\Bbb O}(s,b)+\delta^+_{
f}(s,b)+ \delta^-_{\omega}(s,b)
\ee
where $\delta^+_{\Bbb P}(s,b)$ is the pomeron contribution, $\delta^-_{\Bbb
O}(s,b)$ is the odderon\footnote{The odderon is the counterpart of the
pomeron with $C=-1$} contribution, and $\delta^+_{ f}$,
$\delta^-_{\omega}(s,b)$ are contributions of $f$ meson exchange ($C=+1$)
and $\omega$ meson exchange ($C=-1$).

The trajectories of $f$ and $\omega$ are~\cite{gd93}:
\be
\begin{array}{l}
\alpha_f(t) = 0.69+0.84 t \\
\\
\alpha_\omega (t) = 0.47+0.93 t
\end{array}
\ee
and
\be
\begin{array}{l}
\delta^+ (s,b)=[i+tg\frac{\pi (\alpha(0)-1)}{2}]\frac{c}{s_0}
(\frac{s}{s_0})^{\alpha(0)-1}\frac{e^{-\frac{b^2}{\rho^2}}}{4\pi \rho^2} \\
(C = +1)
\\
\delta^- (s,b)=[i+ctg\frac{\pi (\alpha(0)-1)}{2}]\frac{c}{s_0}
(\frac{s}{s_0})^{\alpha(0)-1}\frac{e^{-\frac{b^2}{\rho^2}}}{4\pi \rho^2}\\
(C = -1)
\end{array}
\ee

Where $i+tg\frac{\pi (\alpha(0)-1)}{2}$ and $i+ctg\frac{\pi
(\alpha(0)-1)}{2}$ are signature coefficients for $C = +1$ and $C = -1$
exchange correspondingly.

\section{Fitting procedure}

We explore two cases: with and without odderon contribution. 

\subsection{The Model With the Odderon}\label{sec:model}

The results concerning the trajectories of the pomeron and the odderon are:
\be
\begin{array}{ll}
\alpha_{\Bbb P}(0) -1 =0.11578\pm0.003 &,\; \alpha'_{\Bbb P}(0) =0.27691
\pm 0.00434
\\
\\
\alpha_{\Bbb O} (0) -1 = 0.11578 \pm 0.00711 &,\;\alpha'_{\Bbb O}(0)
=0.27691 \pm 0.00315
\end{array}
\label{eq:results}
\ee
where $\alpha_{\Bbb P}$ and $\alpha_{\Bbb O}$ are the intercepts of the
pomeron and the odderon, $\alpha'_{\Bbb P}$ and $\alpha'_{\Bbb O}$ are
the slopes of the pomeron and the odderon respectively. So the fitting
procedure lead us practically to the scenario of the weakly degenerate
pomeron-odderon trajectory~\cite{PS}.

$\alpha_{\Bbb P},\alpha'_{\Bbb P}$ and $\alpha_{\Bbb O},\alpha'_{\Bbb O}$
were chosen so that the following conditions of unitarity~\cite{unit}
\be
\begin{array}{l}
\alpha_{\Bbb P} (0) \ge \alpha_{\Bbb O}(0) \\
\\
\alpha'_{\Bbb P} (0) \ge \alpha'_{\Bbb O}(0)
\end{array}
\ee
would not be violated.
The other parameters are in the Table.(\ref{tab:1}).

\begin{table}[h]
\caption{Parameters of the model with the odderon. \label{tab:1}}
\begin{center}
\begin{tabular}{|c|c|c|c|}
\hline
$c_{\Bbb P}$ & 44.66025 &$c_{\Bbb O}$& 1.92343\\
$c_f$& 241.69257 & $c_\omega$& 71.20085 \\
$r^2_{\Bbb P}(GeV^{-2})$ & 14.83200 &$r^2_{\Bbb O}(GeV^{-2})$ & 2.18549 \\
$r^2_{f}(GeV^{-2})$ & 5.55410 &$r^2_{\omega}(GeV^{-2})$ & 99.99996 \\
\hline
\end{tabular}
\end{center}
\end{table}

We find out that the odderon contribution is significant when $t\ne 0$ and
it allows us to qualitatevely describe the differnce in the structure of
$\frac{d\sigma}{dt}$ of $\bar p p$ and $pp$
~\ref{fig:difpp},~\ref{fig:difpbarp} while without this contribution we are
able to describe only the slopes of $\frac{d\sigma}{dt}$ when $t=0$.

It is worth noting that the discrepancy in the description of $\rho =
\frac{ReT}{ImT}(s,t=0)$ can be understood if we remember that the $\rho$
data are not directly mesaured by experimentalists, but rather extracted
from
$\frac{d\sigma}{dt}$ experimental data in a model-dependent way.

\subsection{The Model Without the Odderon}

The results concerning the pomeron trajectory is:
\be
\begin{array}{ll}
\alpha_{\Bbb P}(0) -1 =0.12169 \pm 0.00286 &,\; \alpha'_{\Bbb P}(0) =0.20933
\pm 0.001379
\end{array}
\ee
The other parameters are in the Table.(\ref{tab:2}).

\begin{table}[h]
\caption{Parameters of the model without the odderon. \label{tab:2}}
\begin{center}
\begin{tabular}{|c|c|c|c|}
\hline
$c_{\Bbb P}$ & 46.599 & & \\
$c_f$& 199.630 & $c_\omega$& 150.530 \\
$r^2_{\Bbb P}(GeV^{-2})$ & 21.204 &  &  \\
$r^2_{f}(GeV^{-2})$ & 4.872 &$r^2_{\omega}(GeV^{-2})$ & 39.967 \\
\hline
\end{tabular}
\end{center}
\end{table}
The intercept is slightly higher then in case of when the odderon
contributes to the eikonal. As is shown, the $t=0$ data can be described
without presence of the odderon, while the $t\ne0$ data can be described,
though qualitatively only by means of the odderon. Let's remind our basic
assumptions:  

\begin{itemize}
\item Regge trajectories are linear.

\item The $t$ dependence of $\hat \delta (s,t)$ is exponential i.e. $\hat
\delta (s,t) \sim e^{t\rho^2}$.
\end{itemize}

It is not our aim at present stage to obtain the best fit. What really
matters is to fix a set of Regge parameters some of which are considered 
universal (slopes, intercepts of the trajectories) while others (residues)
are not. Universal parameters will be then used for off-shell amplitudes.

\section{Off-Shell Extention Of The Regge-Eikonal Approach}

As it was shown in~\cite{Petrov} the amplitude with one off-mass shell
particle both in initial and final states (two asteriks)is related to the
off-shell eikonals with one and two off-shell particles and on-shell
amplitude in the following way 
\be
T^{**}(s,b)=\delta^{**}(s,b) -
\frac{\delta^*(s,b)\delta^*(s,b)}{\delta(s,b)}+\frac{\delta^*(s,b)\delta^*(s
,b)}{\delta(s,b)\delta(s,b)}T(s,b)
\label{eq:t**1}
\ee

The expansion Eq.~\ref{eq:t**1}
can be illustrated by the following figure:
\begin{figure}[h]
\vskip -6cm
\hskip -2cm {\vbox to 50mm{\hbox to 50mm{\epsfxsize=120mm
\epsffile{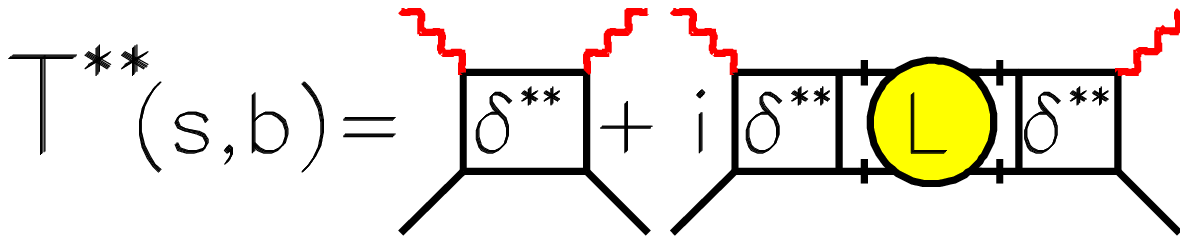}}}}
\vskip 3cm
\end{figure}

where $L=\sum_{n=2}^{\infty}\frac{(2i\delta)^{n-2}}{n!}$.
The case when only one of the particles is off shell can be considered
similarly. One has

\be
T^{*} =\hat\delta^{*} +i\hat\delta^*\circ L\circ\hat\delta
\label{eq:t*}
\ee

\begin{figure}[h]
\vskip -6cm
\hskip -2cm {\vbox to 50mm{\hbox to 50mm{\epsfxsize=120mm
\epsffile{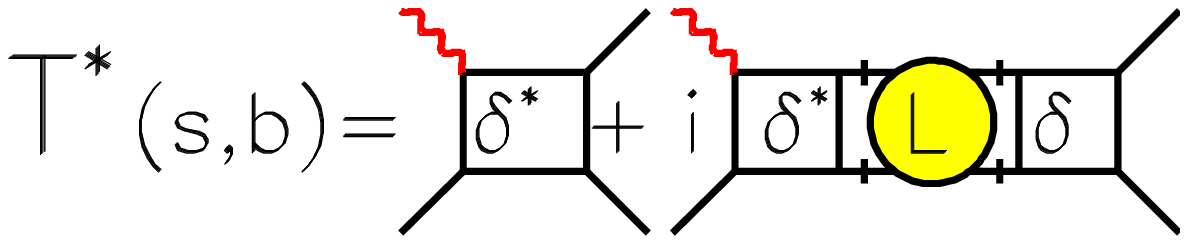}}}}
\vskip 3cm
\end{figure}

or we rewrite it as follows:

\be
T^*(s,b)= \frac{\delta^*(s,b)}{\delta(s,b)}T(s,b)
\label{eq:t*1}
\ee
On the basis of Eq.(~\ref{eq:eikonalb}) we take the
following parametrizations of the off-shell eikonal functions:
\be
\delta^*_{\pm} (s,b)=\xi_{\pm} \frac{c_*(Q^2)}{s_0+Q^2}
\Big(\frac{s+Q^2}{s_0+Q^2}\Big)^{\alpha(0)-1}\frac{e^{-\frac{b^2}{\rho_*^2}}
}{4\pi
\rho_*^2}
\label{eq:offshelleik*}
\ee
where
\be
\rho_*^2 = 4\alpha'(0) ln\frac{s+Q^2}{s_0+Q^2}+r_N^2+r_*^2(Q^2)
\label{eq:offshellrad*}
\ee
and 
\be
\delta^{**}_{\pm} (s,b)=\xi_{\pm} \frac{c_{**}(Q^2)}{s_0+Q^2}
\Big(\frac{s+Q^2}{s_0+Q^2}\Big)^{\alpha(0)-1}\frac{e^{-\frac{b^2}{\rho_{**}^
2}}}{4
\pi \rho_{**}^2}
\label{eq:offshelleik**}
\ee
where
\be
\rho_{**}^2 = 4\alpha'(0) ln\frac{s+Q^2}{s_0+Q^2}+r_N^2+r_{**}^2(Q^2)
\label{eq:offshellrad**}
\ee
Coefficients $c_*(Q^2)$, $c_{**}(Q^2)$ are supposed to weakly (not in a
powerlike way)depend on $Q^2$.

\subsection{Total cross section}
If we consider only the Pomeron exchange (in case of high energies, the main
contribution to the scattering amplitude is due to the Pomeron) and set
signature coefficient $\xi_+$ to be equal unity, then we have the
following series for the total cross section ($\sigma^{**}_{tot} =
\frac{1}{s} ImT^{**}(s,t=0)$):
\be
\sigma^{**}_{tot}=\frac{c_{**}(Q^2)}{Q^2+s_o}\Big(\frac{1}{x}\Big)^\Delta - 
\frac{c_{*}^2(\frac{1}{x})^{2\Delta}}{2 \pi
(Q^2+s_o)^2\rho_*^2}\sum_{n=0}^{\infty}
\frac{\Big(-\frac{c(\frac{s}{s_o})^{\Delta}}{2 \pi
s_o\rho^2}\Big)^n}{(n+2)!}\cdot \frac{\rho^2}{2\rho^2+n\rho_*^2}
\label{eq:totalcrs}
\ee
here we used the folowing relation
$\frac{s+Q^2}{s_0+Q^2}\simeq\frac{s+Q^2-m_N^2}{Q^2}=\frac{1}{x}$ (if $s \gg
Q^2$ and (or)$Q^2\gg s_0 (Q^2\gg m_N^2)$)
Now we can derive the behaviour of the total cross section in different
kinematic limits:
\begin{itemize}
\item {\bf Regge Regime} 
\be
\sigma^{**}_{tot} \rightarrow \frac{(s/Q^2)^\Delta}{Q^2}\Big[c_{**}-
\frac{c_*^2}{c}\Big(\frac{s_0}{Q^2}\Big)^{1+\Delta}\frac{\rho^2}{\rho_*^2}
\Big]
\label{eq:totalregge}
\ee
\item {\bf Bjorken Regime}
\be
\sigma^{**}_{tot} \rightarrow
\frac{c_{**}(Q^2)}{Q^2}\Big(\frac{1}{x}\Big)^\Delta
-\frac{c_*^2}{2c}\cdot \frac{1}{Q^2}\cdot \Big(\frac{1}{x}\Big)^\Delta
\cdot 
\Big(\frac{s_o}{Q^2}\Big)^{1+\Delta} \cdot \frac{ln \frac{Q^2(1-x)}{s_o
x}}{ln\frac{1}{x}}
\label{eq:totalbjorken}
\ee
\end{itemize}
As we see total cross-section posesses a powerlike behaviour in the Regge
limit, but this is not a violation of unitarity, as the Froissart-Martin
bound~\cite{fm} cannot be proven for this case and if we put all particles
on the mass shell, then we restore the `normal' logarithmic asymptotical
behaviour $\sigma\sim\ln^2\frac{s}{s_0}$. 
In the Bjorken limit we have strong (powerlike) violation of scaling in the
second term. 
\subsection{Elastic cross-section}
For the elastic cross section we have the following
expression:
\be
\sigma^{*}_{el} = 4 \pi \int_{0}^{\infty} db^2 \Big\vert
\frac{\delta^*}{\delta} T(s,b)\Big\vert^2
\label{eq:elasticcrs}
\ee
As far as $q'^2=\mu^2$, where $\mu$ is the mass of the produced particle, it
is
natural to set $s_0 = \mu^2$ and now we can derive the following relations:
\begin{itemize}
\item {\bf Regge Regime} 
\be
\sigma^{*}_{el} \rightarrow 16\pi \alpha'(0)\Delta\Big(\frac{c_*}{c}\Big)^2
\Big(\frac{\mu^2}{Q^2}\Big)^{2+2\Delta}(ln\frac{s}{\mu^2})^2
\label{eq:elasticregge}
\ee
\item {\bf Bjorken Regime}
\be
\sigma^{*}_{el} \rightarrow 8 \pi \alpha'(0) \Big(\frac{c_*}{c}\Big)^2
\Big(\frac{\mu^2}{Q^2}\Big)^{2+2\Delta}\frac{(ln(Q^2/x))^2}{ln(1/x)}
\label{eq:elasticbjorken}
\ee
\end{itemize}
As we can easily realize
\be
\frac{\sigma^{*}_{el}}{\sigma^{**}_{tot}} \rightarrow 0
\label{eq:eltot}
\ee
when $s\rightarrow \infty$ and $s\gg Q^2$ (compare with~\ref{eq:ratio})
As to the $Q^2$ behaviour, then as far as we have $\Delta =
0.116\pm0.003$ (see Eq.(~\ref{eq:results})), we obtain the following
dependence:
\be
\sigma^{*}_{el}\sim\frac{1}{(Q^2)^{2+2\Delta}}=(Q^2)^{-2.232\pm0.006}
\ee
This formula is in close agreement with the results of fitting
the $Q^2$ dependence in~\cite{H1}, where the power of $Q^{-2}$ is
measured as $2.24$.

\section{Vector meson photoproduction}

For vector meson photoproduction $\gamma^* p \rightarrow V p$, where
$V=\rho_o,\omega, \Phi, J/\Psi ,...$
 
\begin{figure}[hb]
 \vskip -6cm
 \hskip 0cm {\vbox to 30mm{\hbox to 30mm{\epsfxsize=120mm
 \epsffile{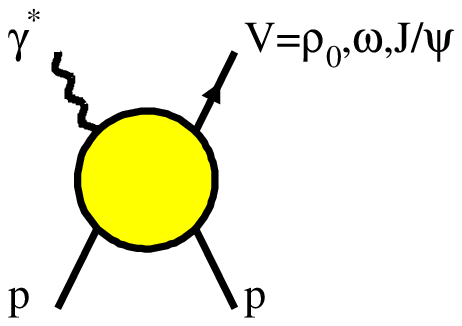}}}}
\vskip 5cm 
\end{figure}

\noindent
we have the following formula for the amplitude:
\be
T^*(s,b)= \frac{\delta^*(s,b)}{\delta(s,b)}T(s,b)
\label{eq:vector}
\ee
here we assume that $T_{\gamma^* p \rightarrow V p}\sim T_{V^* p
\rightarrow
V p}$ on the basis of vector dominance~\cite{vectordominance}.
As far as in this process reggeons with $C=-1$ do not contribute, we have
the following contributions for the eikonal function:
\be
\delta(s,b) = \delta^+_{\Bbb P}(s,b)+ \delta^+_{ f}(s,b)
\ee
with ($s\equiv W^2$)
\be
\delta_\pm (W,b)=\xi_\pm \frac{c}{W^2-\mu^2-m_p^2}
\Bigg(\frac{W^2-\mu^2-m_p^2}{W_0^2-\mu^2-m_p^2}\Bigg)^{\alpha(0)}\frac{e^{-
\frac{b^2}{\rho^2}}}{4\pi \rho^2}
\label{eq:vectoreik}
\ee
\be
\rho^2(W) = 4 \alpha'(0) ln\frac{W^2-\mu^2-m_p^2}{W_0^2-\mu^2-m_p^2}+r_p^2
\label{eq:vectorrad}
\ee
and 
\be
\delta_\pm^* (W,b)=\xi_\pm \frac{c_*(Q^2)}{W^2+Q^2-m_p^2}
\Bigg(\frac{W^2+Q^2-m_p^2}{W_0^2+Q^2-m_p^2}\Bigg)^{\alpha(0)}\frac{e^{-\frac
{b^2}{\rho_*^2}}}{4\pi \rho_*^2}
\label{eq:vectoreik*}
\ee
\be
\rho_*^2(W) = 4 \alpha'(0)
ln\frac{W^2+Q^2-m_p^2}{W_0^2+Q^2-m_p^2}+r_p^2+r_*^2(Q^2)
\label{eq:vectorrad*}
\ee
As far as Regge Eikonal approach does not constrain the $Q^2$ dependence
of residues $c_*(Q^2)$ and radii $r_*^2(Q^2)$ we assume that the main $Q^2$
dependence is comprised in the other terms of our formulae and residues
slowly depend on $Q^2$ (they posses logarifmic behaviour)and radii depend
on $Q^2$ as $\sim \frac{1}{Q^2}$.
Eventually we have the following parametrizations:
\be
c_*(Q^2)=c+c_1
ln^3(\frac{Q_o^2+Q^2}{Q_o^2})
\ee
and
\be
r_*^2(Q^2)=r^2+\frac{r^2_1}{\frac{Q_0^2+Q^2}{Q_0^2}}
\ee
where $Q_0^2= 1.0 GeV^2$.

The results of fitting the data to these formulae are in Fig.1.

\begin{figure}[ht]
\vskip -0.5cm
\begin{center}
 \hskip -6cm {\vbox to 50mm{\hbox to 50mm{\epsfxsize=50mm
 \epsffile{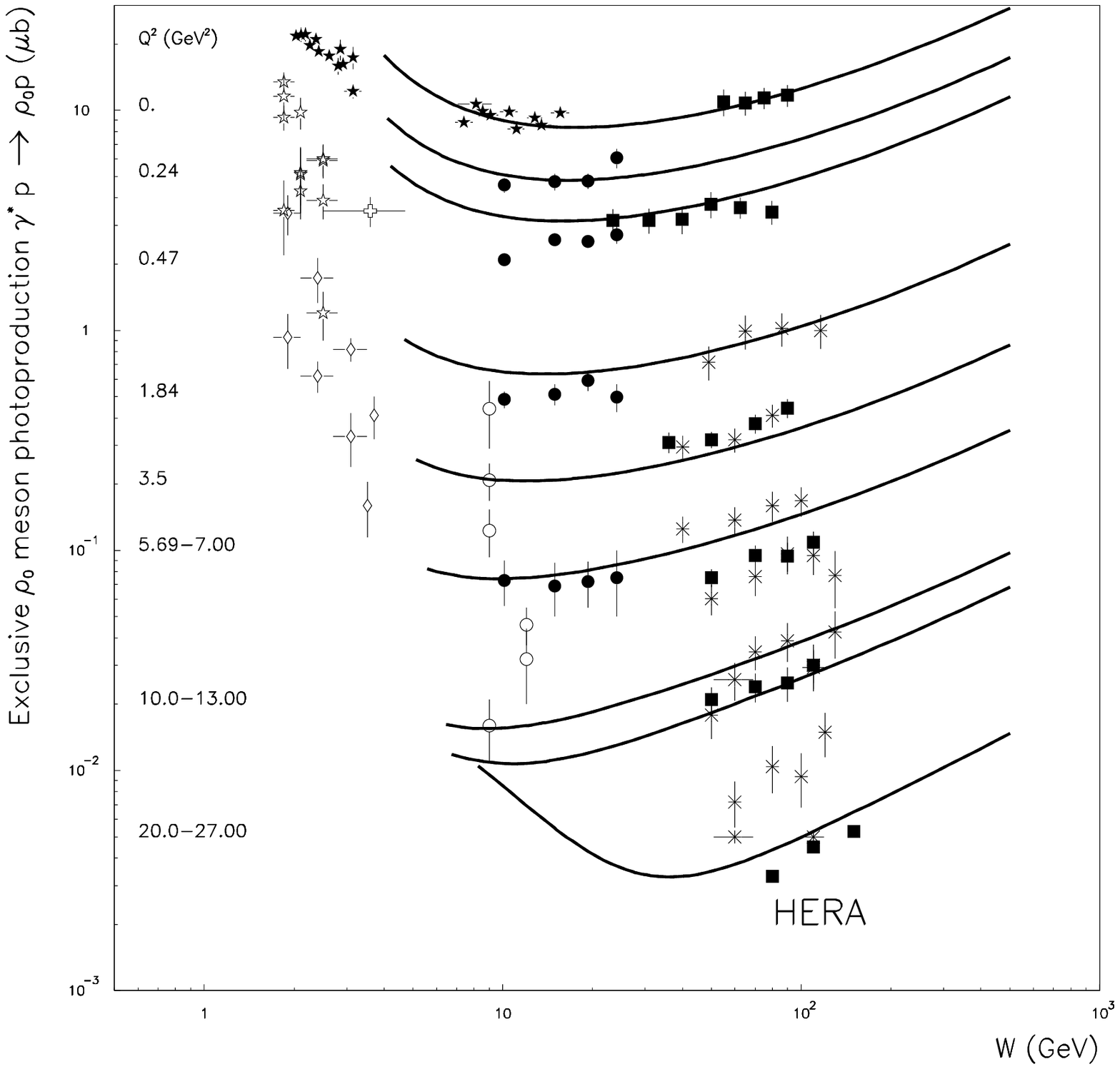}}}}
 \vskip -5cm
 \hskip 6cm {\vbox to 50mm{\hbox to 50mm{\epsfxsize=50mm
 \epsffile{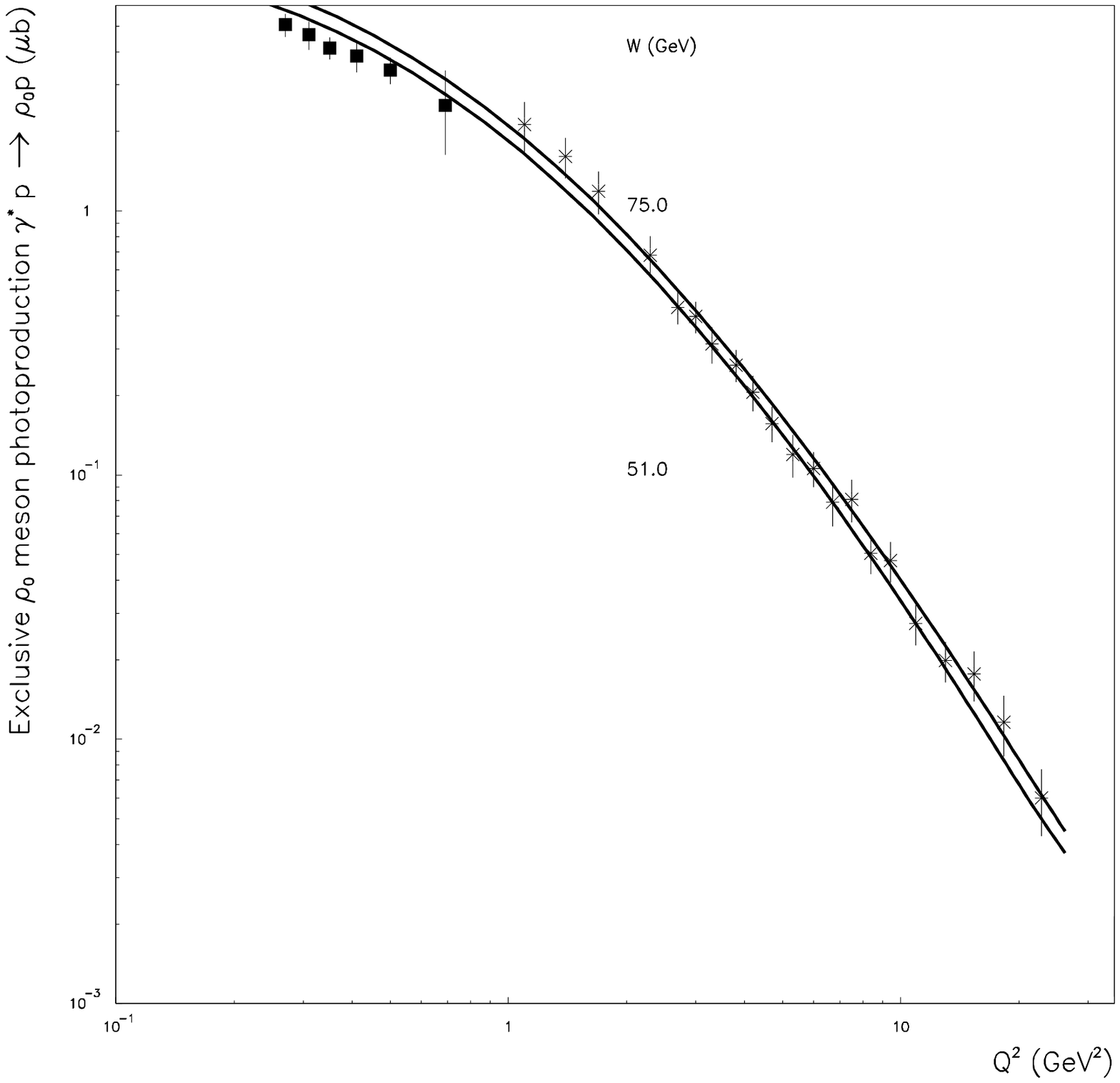}}}}
\end{center} 
\vskip 1.5cm
\label{fig:rho0}
\caption{$\rho_0$ meson exclusive photoproduction cross-section}
\vskip 1cm
\end{figure}
\vskip -1cm
\begin{figure}[ht]
\begin{center}
 \hskip -6cm {\vbox to 50mm{\hbox to 50mm{
 \epsfxsize=50mm\epsffile{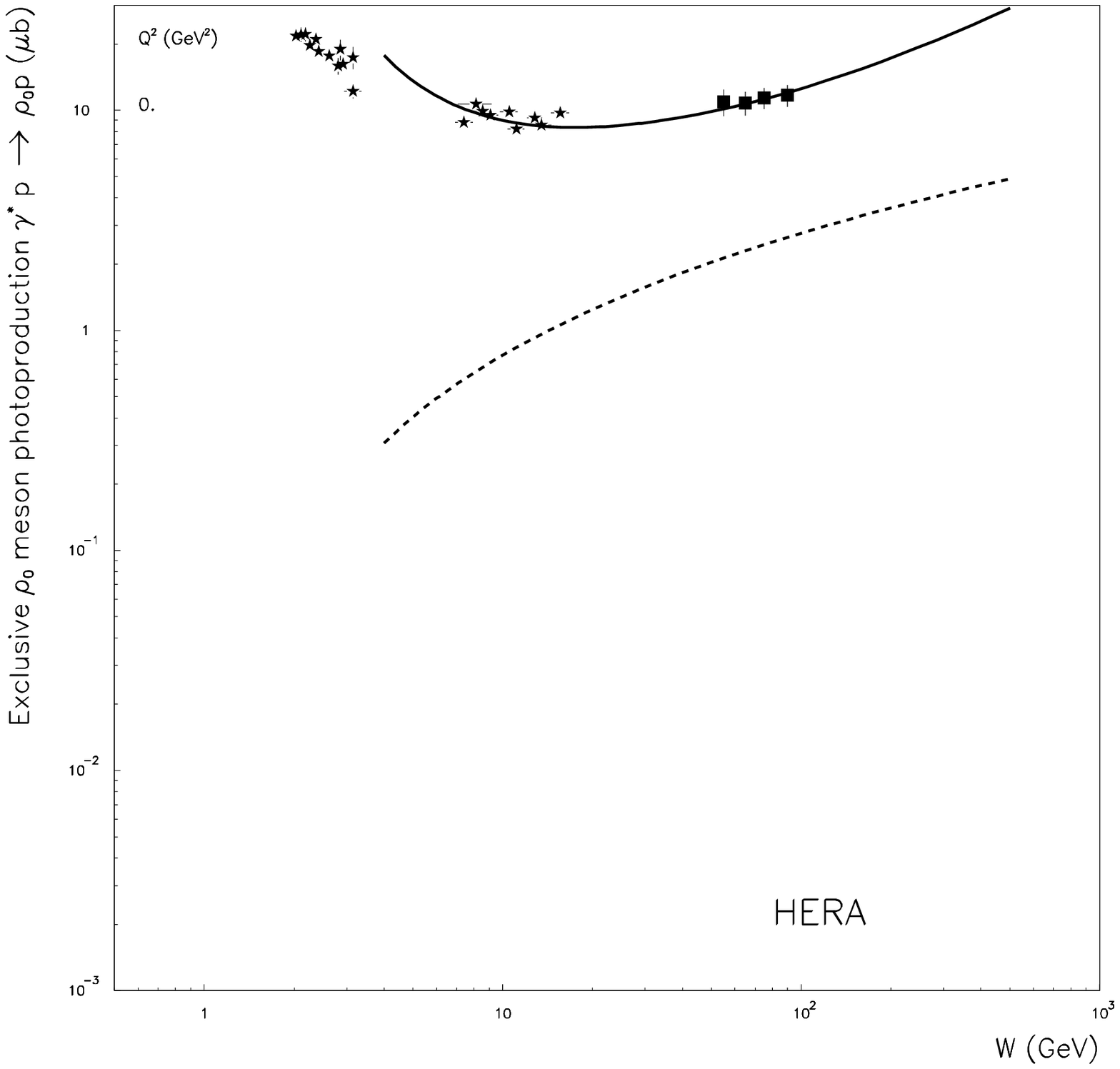}}}}
 \vskip -5cm
 \hskip 6cm {\vbox to 50mm{\hbox to 50mm{
 \epsfxsize=50mm\epsffile{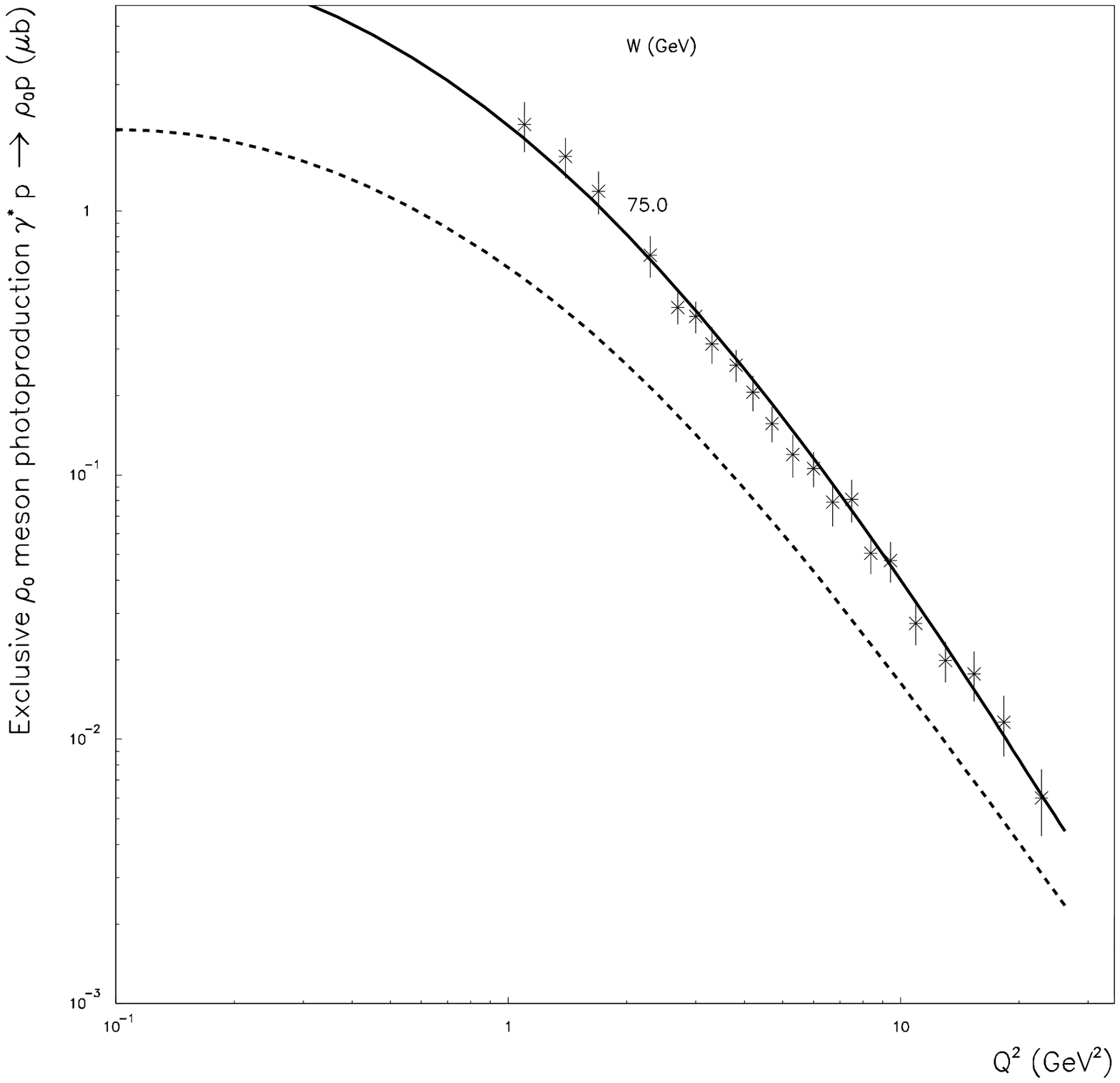}}}}
\end{center} 
\vskip 1.5cm
\label{fig:rho10}
\caption{$\rho_0$ meson exclusive photoproduction cross-section (solid
line)Eq. \ref{eq:elasticregge} in comparison with its asymptotic behaviour
(dashed
line) Eq.~\ref{eq:elasticbjorken}}
\end{figure}
As one can see from Fig.2, in the region of energies that are available the
asymptotic behaviour of exclusive vector meson photoproduction does not
take place and therefore one should take into concideration both Pomeron and
$f$ meson exchanges. 

For fitting we used the same intercept and slope of the pomeron that have
been obtained in the Section~\ref{sec:model}, so that the pomeron in our
model is unique and the same for all processes where it takes part.

\section{Conclusions}

As is seen from Fig.1 and Fig.2, the exclusive vector
meson production may be described in the framework of extended Regge-eikonal
approach and one does not need `hard' pomerons with a high
intercept in order to do it. The rise of cross sections when $Q^2$ is high
is a transitory phenomenon due to effective delay of asymptotic behaviour
which is observed for low $Q^2$~\cite{Martynov} or ~\cite{Prokudin}. The
$Q^2$-dependence is also  described fairly well. 




\newpage

\begin{figure}[t]
\begin{center}
{\bf Results of fitting without the odderon contribution} \\
Hollow dots are $pp$ data and full dots are $\bar p p$ data. The dotted
($pp$) and solid ($\bar p p$) curves correspond to the model without the
odderon contribution. The shadowed area corresponds to the region available
by the uncertainty in the fitting parameters.

\end{center}
\vskip 1.5cm
\begin{minipage}{58mm}
\vskip -2.cm
{\vbox to 55mm{\hbox to 55mm{\epsfxsize=55mm
\epsffile{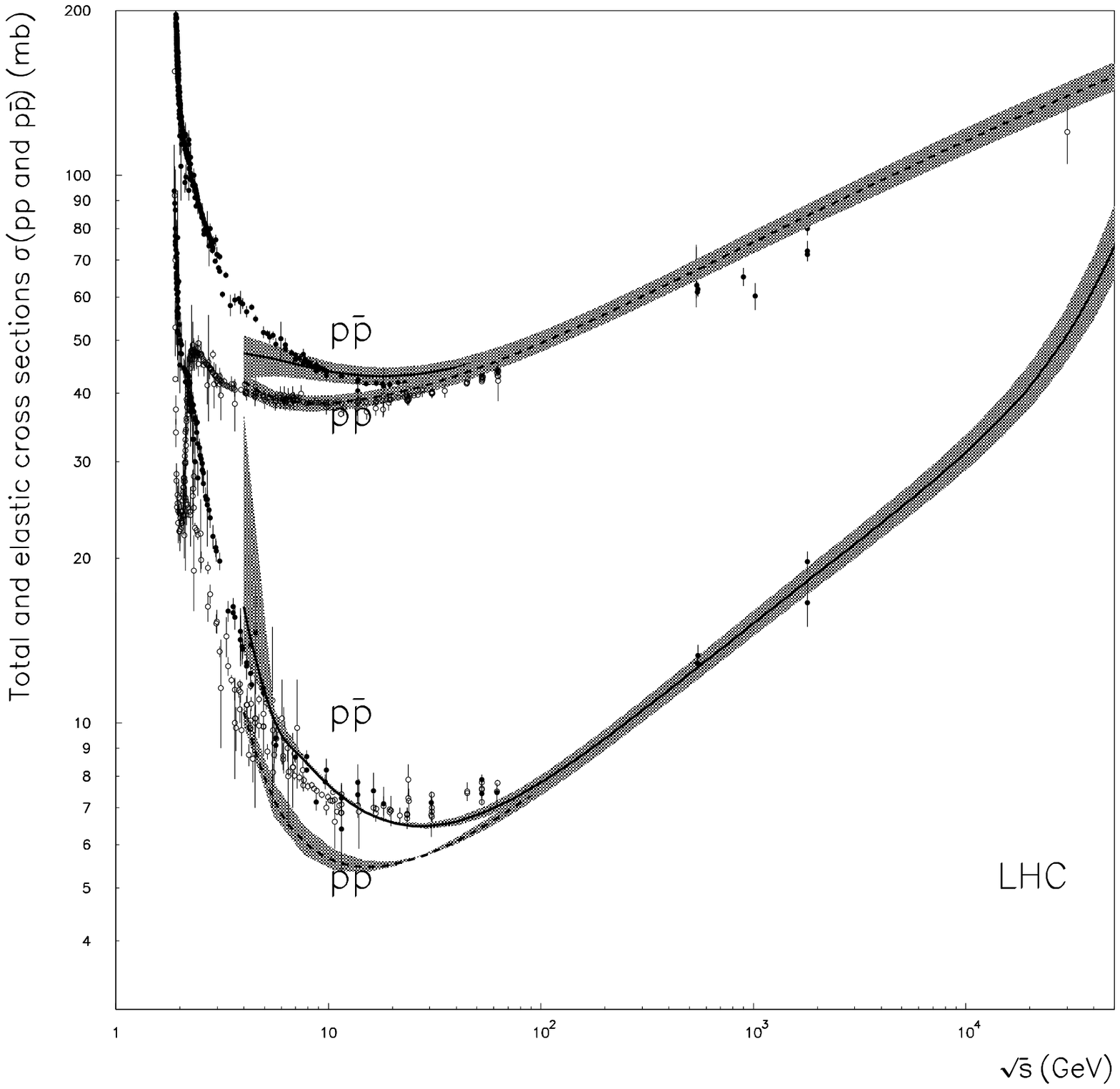}}}}
\vskip 0.5cm
\caption{Fitted total and elstic cross sections.
\label{fig:totno}}
\end{minipage}
\hskip 0.1cm
\begin{minipage}{58mm}
\vskip -2.cm
{\vbox to 55mm{\hbox to 55mm{\epsfxsize=55mm
\epsffile{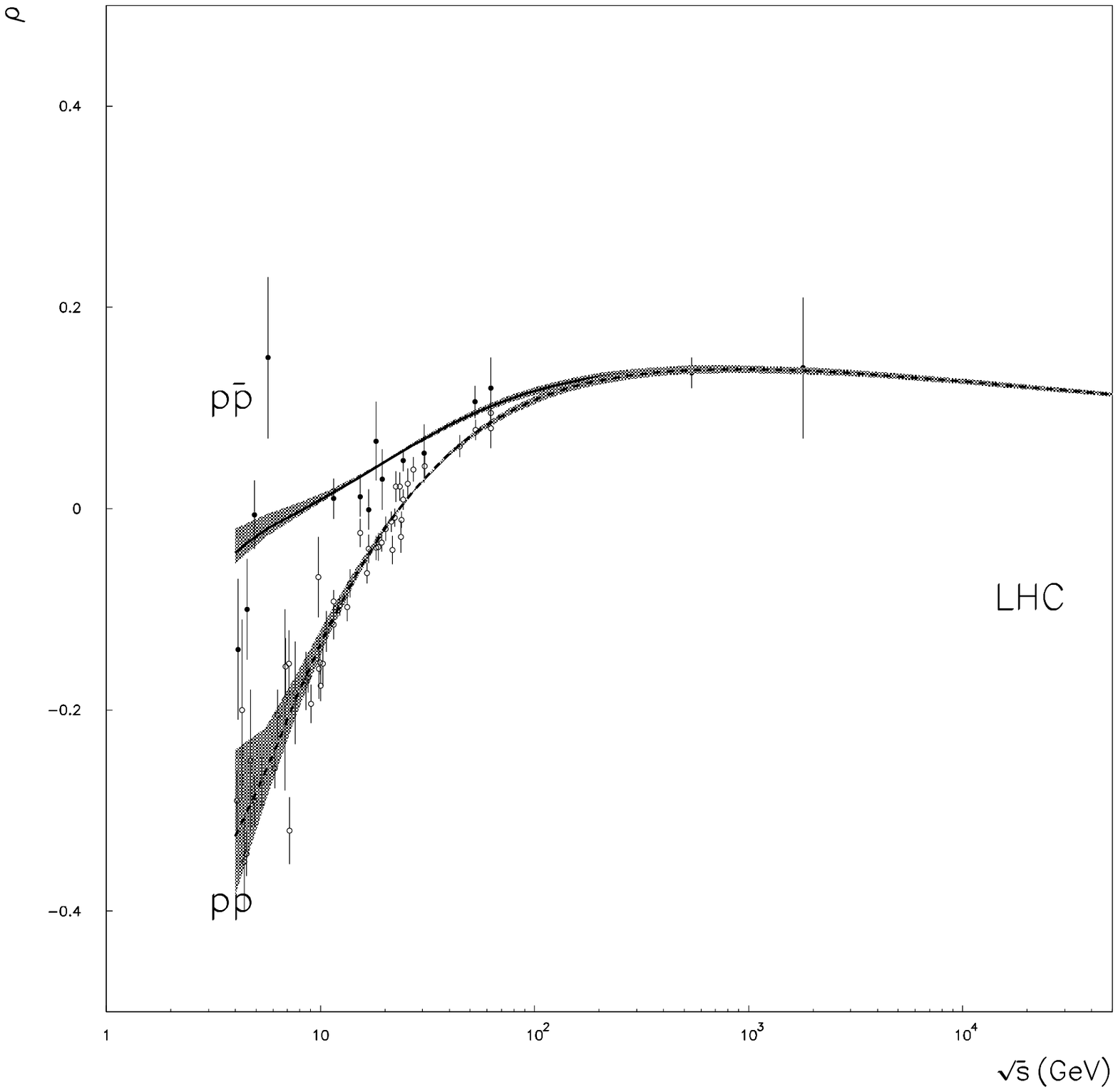}}}}
\vskip 0.5cm
\caption{Fitted ratios of real to imaginary forward amplitudes.
\label{fig:rhono}}
\end{minipage}
\vskip 1.5cm
\begin{minipage}{58mm}
\vskip -2.cm
{\vbox to 55mm{\hbox to 55mm{\epsfxsize=55mm
\epsffile{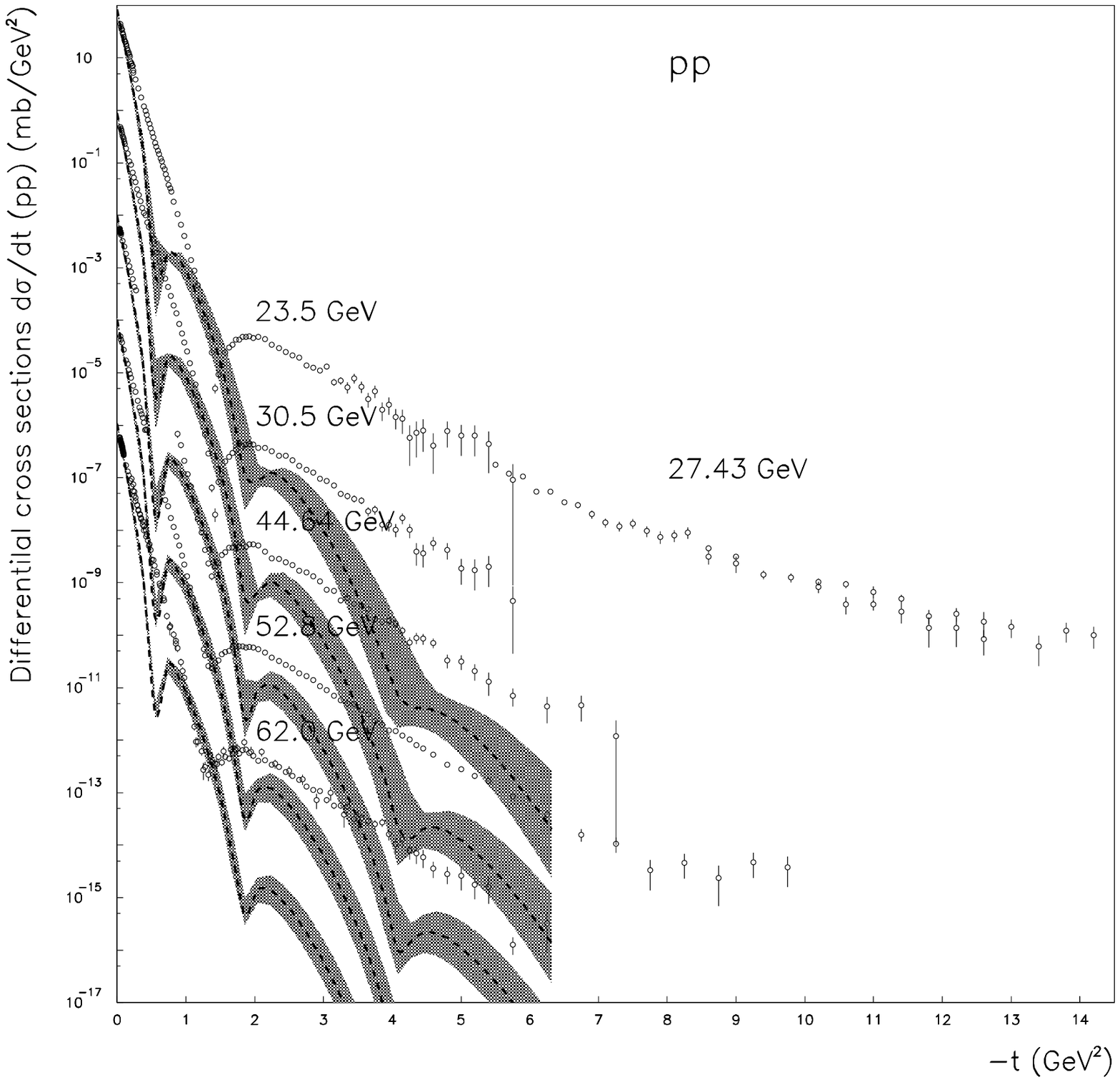}}}}
\vskip 0.5cm
\caption{Fitted differential cross sections ($pp$).
\label{fig:difppno}}
\end{minipage}
\hskip 0.01cm
\begin{minipage}{58mm}
\vskip -2.cm
{\vbox to 55mm{\hbox to 55mm{\epsfxsize=55mm
\epsffile{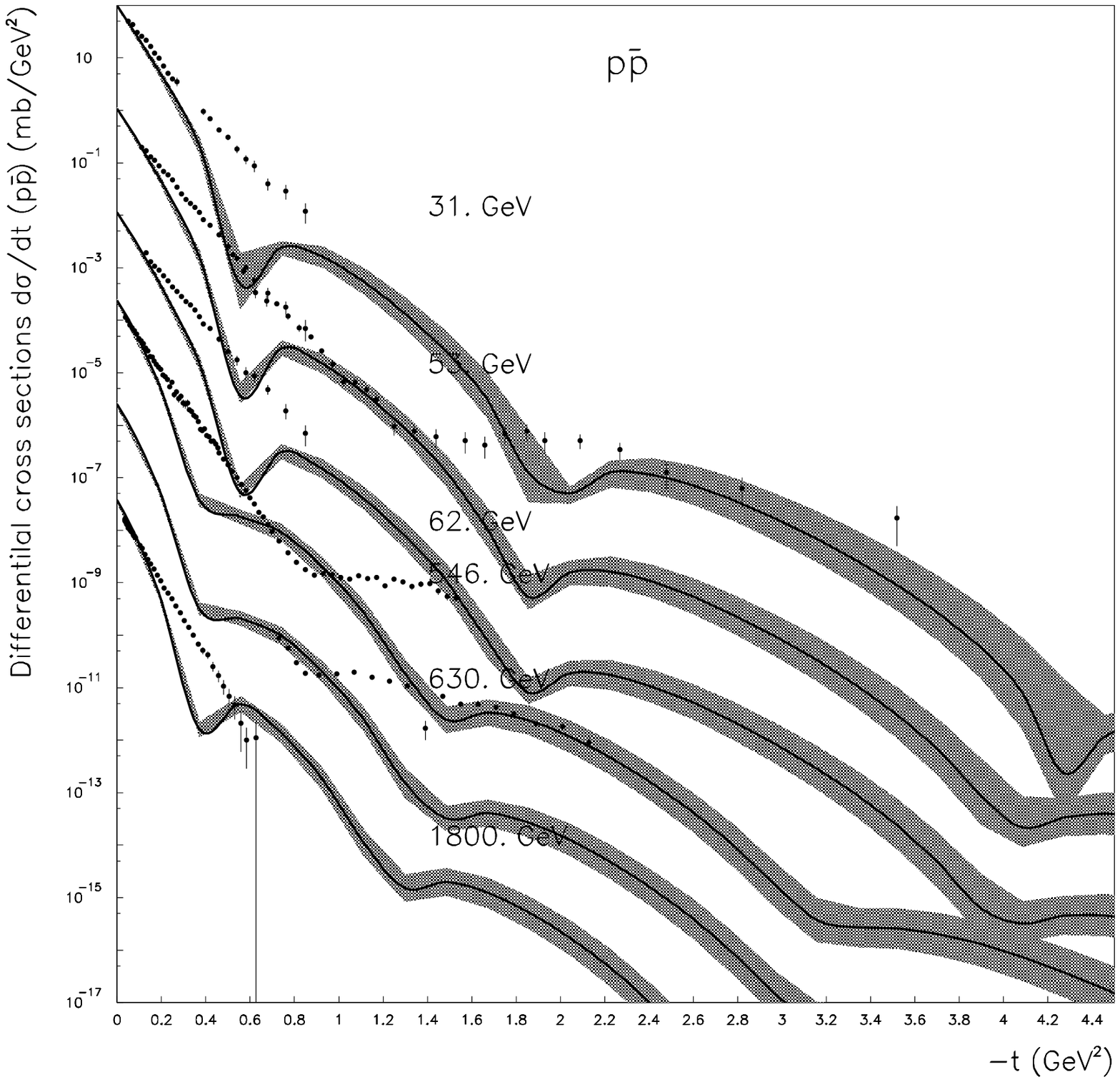}}}}
\vskip 0.5cm
\caption{Fitted differential cross sections ($\bar p p$).
\label{fig:difpbarpno}}
\end{minipage}
\end{figure}
\newpage
\begin{figure}[t]
\vskip -4.7cm
\begin{center}
{\bf Results of fitting with the odderon contribution} \\
Hollow dots are $pp$ data and full dots are $\bar p p$ data. The dotted
($pp$) and solid ($\bar p p$) curves correspond to the model with the
odderon contribution. The shadowed area corresponds to the region available
by the uncertainty in the fitting parameters.
\end{center}
\vskip 1.5cm
\begin{minipage}{58mm}
\vskip -2cm
{\vbox to 55mm{\hbox to 55mm{\epsfxsize=55mm
\epsffile{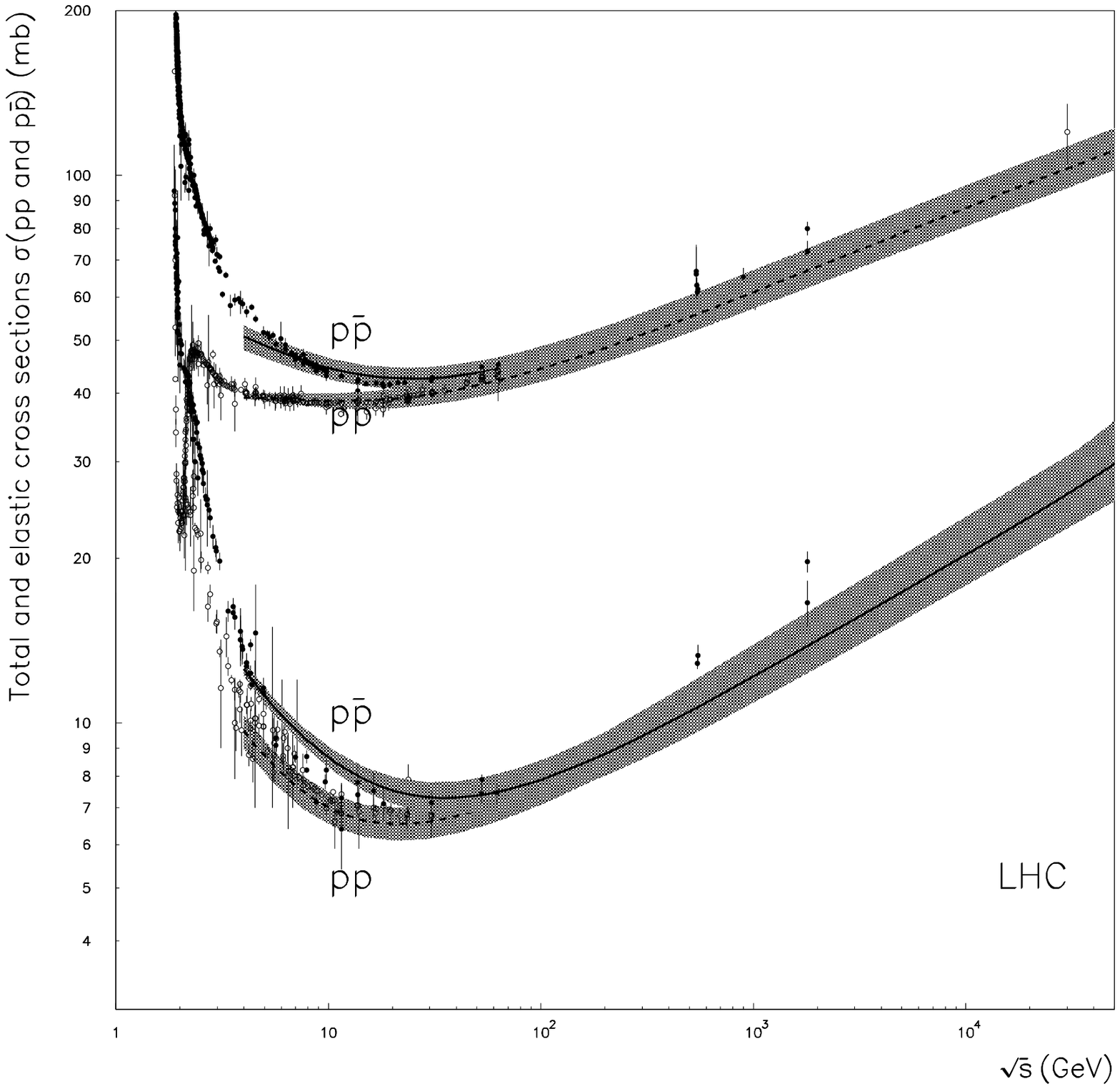}}}}
\vskip 0.5cm
\caption{Fitted total and elstic cross sections.
\label{fig:tot}}
\end{minipage}
\hskip 0.1cm
\begin{minipage}{58mm}
\vskip -2cm
{\vbox to 55mm{\hbox to 55mm{\epsfxsize=55mm
\epsffile{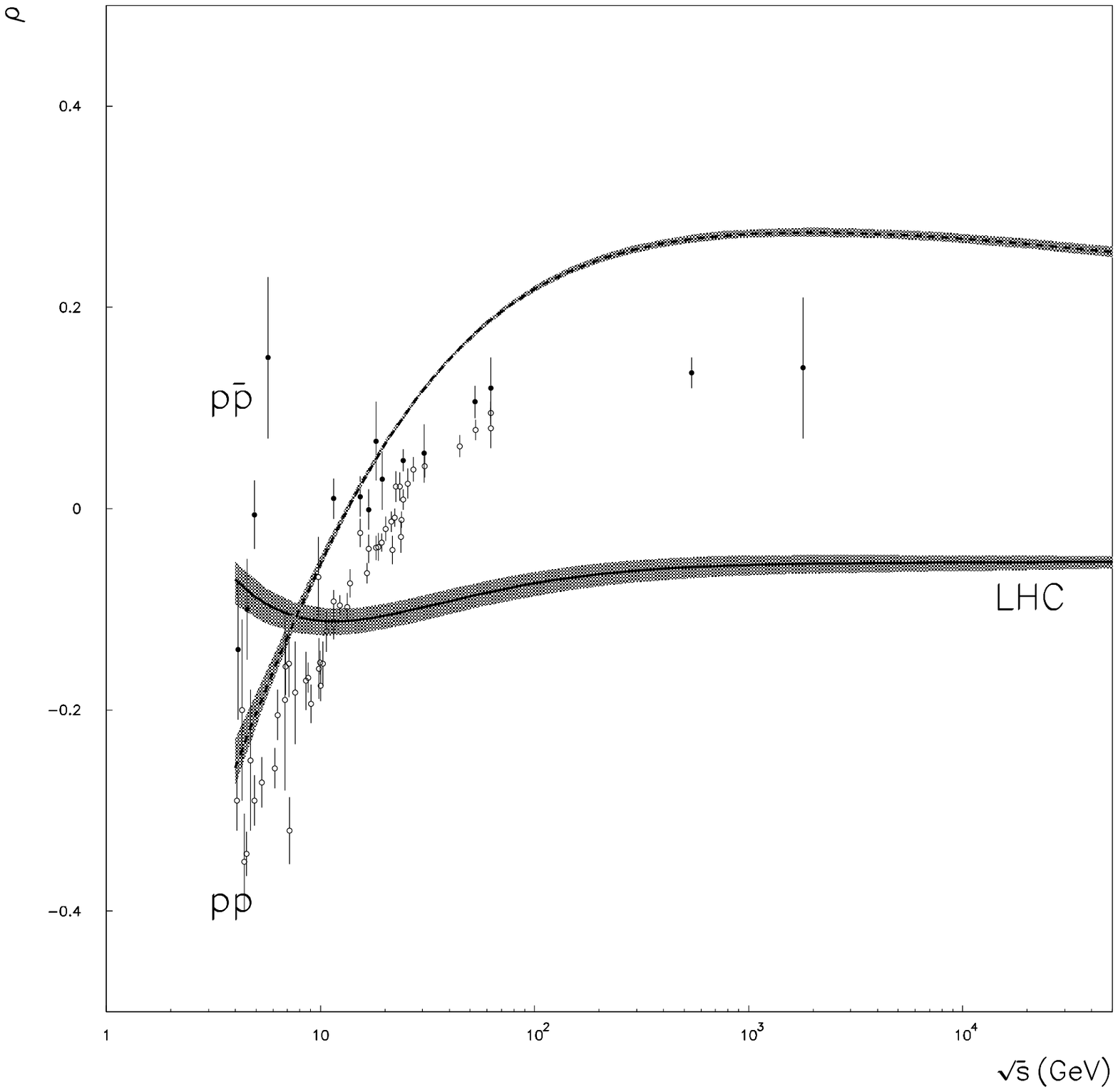}}}}
\vskip 0.5cm
\caption{Fitted ratios of real to imaginary forward amplitudes.
\label{fig:rho}}
\end{minipage}
\vskip 1.5cm
\begin{minipage}{58mm}
\vskip -2cm
{\vbox to 55mm{\hbox to 55mm{\epsfxsize=55mm
\epsffile{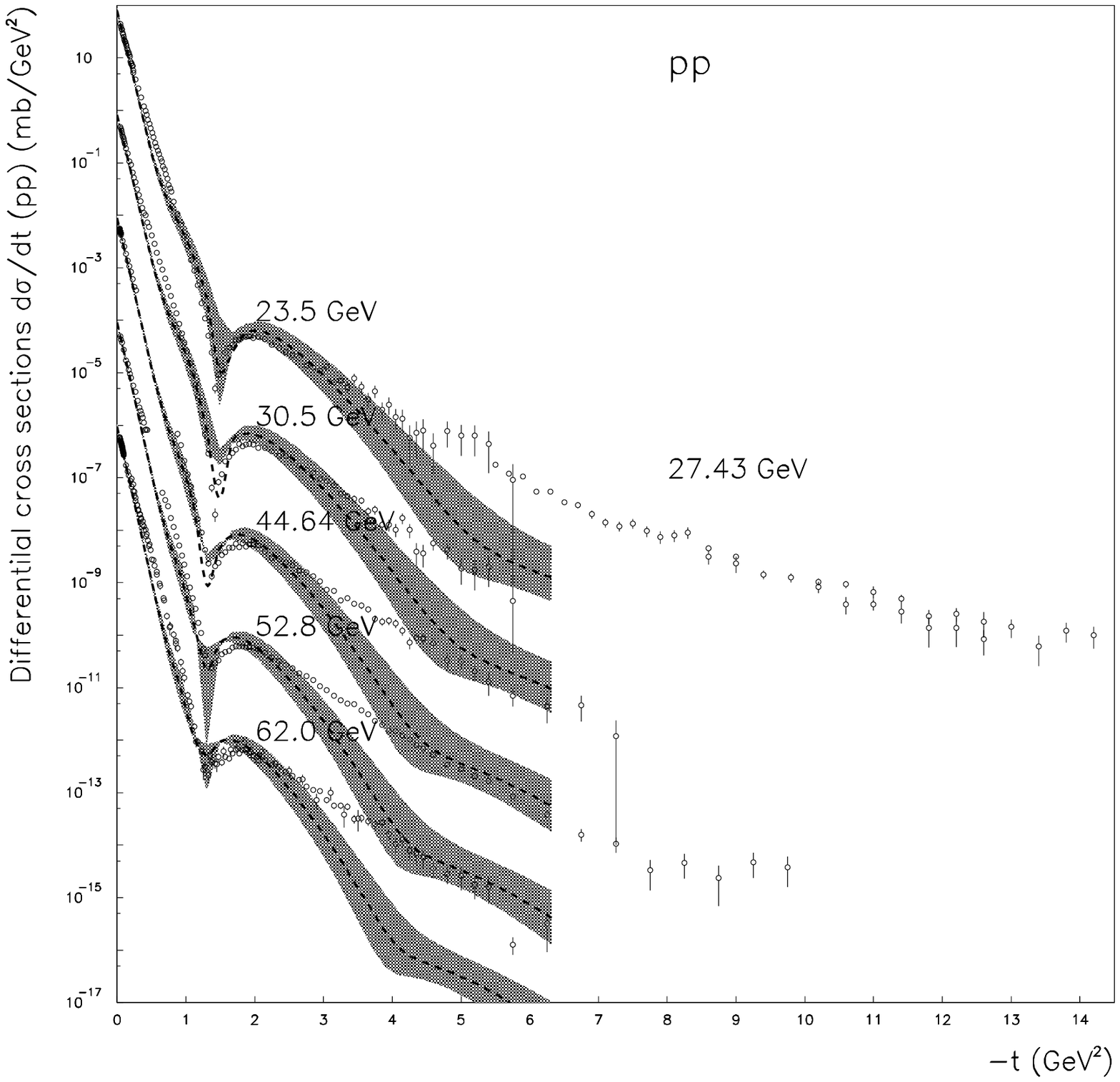}}}}
\vskip 0.5cm
\caption{Fitted differential cross sections ($pp$).
\label{fig:difpp}}
\end{minipage}
\hskip 0.01cm
\begin{minipage}{58mm}
\vskip -2cm
{\vbox to 55mm{\hbox to 55mm{\epsfxsize=55mm
\epsffile{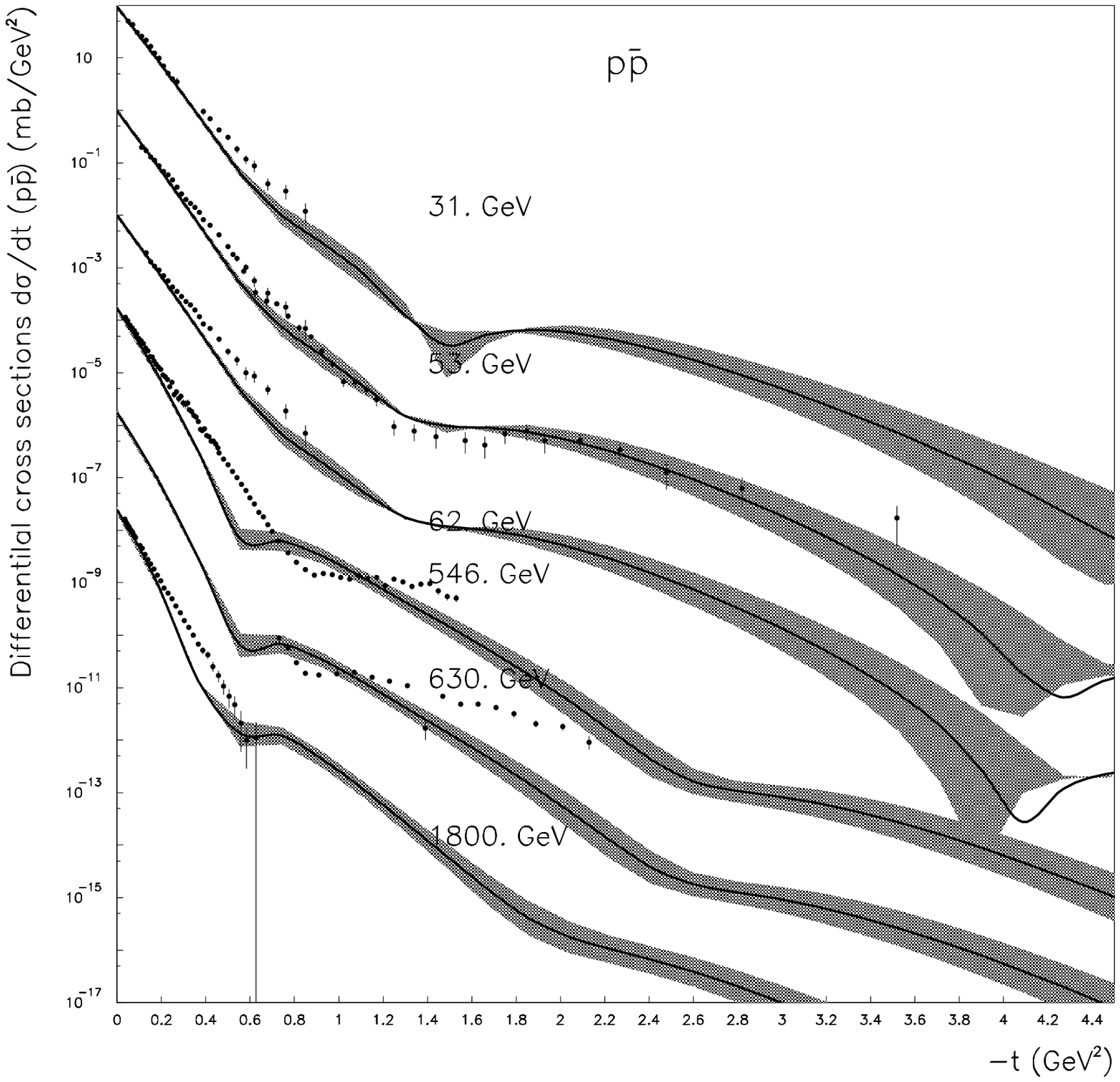}}}}
\vskip 0.5cm
\caption{Fitted differential cross sections ($\bar p p$).
\label{fig:difpbarp}}
\end{minipage}
\end{figure}

\end{document}